\documentstyle[12pt]{article}

\catcode`@=11

\def\eqnarray{\stepcounter{equation}\let\@currentlabel=\theequation
\global\@eqnswtrue
\global\@eqcnt\z@\tabskip\@centering\let\\=\@eqncr
$$\halign to \displaywidth\bgroup\@eqnsel\hskip\@centering
  $\displaystyle\tabskip\z@{##}$&\global\@eqcnt\@ne
  \hskip 3pt \hfil${##}$\hfil
  &\global\@eqcnt\tw@ \hskip 3pt $\displaystyle\tabskip\z@{##}$\hfil
  \tabskip\@centering&\llap{##}\tabskip\z@\cr}

\def\@begintheorem#1#2{\trivlist
      \item[\hskip \labelsep{\bf #1\ #2.}] \sl}
\def\@opargbegintheorem#1#2#3{\trivlist
      \item[\hskip \labelsep{\bf #1\ #2\ (#3).}] \sl}

\catcode`@=12

\newcommand{\Bbb}{\bf}

\newenvironment{proof}{\trivlist\item[\hskip\labelsep{\it Proof.}]
 \rm}{\hfill\framebox[.6em]{\rule{0em}{1ex}}\endtrivlist}

\newtheorem{theorem}{Theorem}
\newtheorem{proposition}[theorem]{Proposition}
\newtheorem{lemma}[theorem]{Lemma}

\newcommand{\real}{{\Bbb R}}
\newcommand{\integer}{{\Bbb Z}}

\newcommand{\APC}{{\it APC}}
\newcommand{\BAL}{{\it BAL}}

\newcommand{\tnb}[1]{{\lowercase{\bf #1}}}
\newcommand{\TTT}{\tnb{T}}
\newcommand{\NNN}{\tnb{N}}
\newcommand{\BBB}{\tnb{B}}
\newcommand{\XXX}{\tnb{X}}
\newcommand{\YYY}{\tnb{Y}}
\newcommand{\ZZZ}{\tnb{Z}}

\newcommand{\nb}[1]{#1}
\newcommand{\NN}{\nb{N}}
\newcommand{\BB}{\nb{B}}
\newcommand{\XX}{\nb{X}}
\newcommand{\YY}{\nb{Y}}
\newcommand{\ZZ}{\nb{Z}}
\newcommand{\UU}{\nb{U}}

\newcommand{\metric}[2]{\langle#1,\:#2\rangle}
\newcommand{\ders}{\partial_s}
\newcommand{\ints}{\partial_s^{-1}}
\newcommand{\Ders}{\nabla_{\!s}\,}
\newcommand{\vari}[1]{\delta_{#1}\,}
\newcommand{\Vari}[1]{\nabla_{\!#1}\,}
\newcommand{\VARI}[1]{\tilde\nabla{}_{\!#1}\,}

\newcommand{\exprn}[1]{{\cal E}_{#1}}
\newcommand{\alge}[1]{{\cal A}_{#1}}
\newcommand{\func}{{\cal F}}
\newcommand{\vect}{{\cal X}}
\newcommand{\tang}[1]{{\cal T}_{#1}}
\newcommand{\form}[1]{{\cal D}^{#1}}

\newcommand{\inner}[2]{\langle#1,\:#2\rangle_\bot}

\newcommand{\injec}{\wp}
\newcommand{\surj}{\varphi}

\newcommand{\ed}{{\it d}}
\newcommand{\ip}[1]{\iota_{#1}}
\newcommand{\Ld}[1]{{\cal L}_{#1}}
\newcommand{\cp}[2]{[#1,\:#2]}
\newcommand{\Sb}[2]{[#1,\:#2]}

\newcommand{\Riem}{g}
\newcommand{\Riemann}[2]{\Riem(#1,\:#2)}
\newcommand{\Poisson}[2]{\{#1,\:#2\}}

\newcommand{\grad}{\mathop{\rm grad}}
\newcommand{\cycle}{{\rm cycle}}
\newcommand{\length}{\ell}

\begin{document}

\def\thefootnote{\fnsymbol{footnote}}

\begin{flushright}
Preprint OCU-164, hep-th/9611???
\end{flushright}
\bigskip
\bigskip
\begin{center}
 \LARGE    {\bf Differential Geometry of\\ the Vortex Filament Equation}
 \bigskip\medskip\\
 \large    Yukinori Yasui\footnote
 {e-mail address: yasui@sci.osaka-cu.ac.jp}\\
 \normalsize
 \rm       and\\
 \large    Norihito Sasaki\footnote
 {e-mail address: norih@sci.osaka-cu.ac.jp}
 \medskip\\
 \normalsize
 \it       Department of Physics, Osaka City University,
           Sumiyoshi-ku, Osaka 558, Japan
 \bigskip\\
 \rm       November 1996
\end{center}
\bigskip\bigskip
\bigskip\bigskip

\noindent
{\bf Abstract.}
Differential calculus on the space of asymptotically linear curves
is developed. The calculus is applied to the vortex filament equation
in its Hamiltonian description. The recursion operator generating
the infinite sequence of commuting flows is shown to be hereditary.
The system is shown to have a description with a Hamiltonian pair.
Master symmetries are found and are applied to deriving an expression
of the constants of motion in involution. The expression agrees with
the inspection of Langer and Perline.
\bigskip
\bigskip

\bigskip
\medskip


\bigskip\bigskip
\bigskip\bigskip

\section{Introduction}

The vortex filament equation \cite{DaR} is a nonlinear evolution
equation describing the time development of a very thin vortex tube.
The equation is derived from the dynamics of 3-dimensional
incompressible fluid with the local induction approximation
and is written as
\begin{equation}
 \dot\gamma = \kappa \BBB, \label{eq.1}
\end{equation}
where $\gamma$ is the curve of the vortex filament parametrized
by the arclength, dot stands for the differential with respect to
the time, $\kappa$ is the curveature of $\gamma$, and $\BBB$ is
the bi-normal vector along $\gamma$.
It is well-known that the vortex filament equation
(\ref{eq.1}) is closely related to the cubic nonlinear
Schr\"odinger (NLS) equation, and the Hasimoto map provides
a connection between them \cite{Has}.
The NLS equation is an infinite-dimensional
completely integrable Hamiltonian system \cite{F-T}.

Marsden and Weinstein \cite{M-W} constructed a Hamiltonian
description of the vortex filament equation in their study on the
moment map for the action of the unimodular diffeomorphism group of
$\real^3$. Langer and Perline \cite{L-P} introduced the space
$\BAL$ --- the space of balanced asymptotically linear curves (see
Section~\ref{sect-2}) --- as a phase space for the system of
vortex filament, and showed that the Hasimoto map is
a Poisson map from $\BAL$ with the Marsden-Weinstein Poisson
structure to (a certain equivalence class of) the phase space of
the NLS system with the `fourth' Poisson structure.
This result says that the Hasimoto map induces constants of motion
in involution for the vortex filament equation as the pull-back
of those for the NLS system, hence the vortex filament equation can
be understood as a completely integrable system.
Further, Langer and Perline found a recursion operator,
which generates infinite sequence of commuting Hamiltonian
vector fields on $\BAL$.

For some typical integrable Hamiltonian systems,
such as the NLS equation, the
integrability is studied from various aspects and many remarkable
structures are known to exist \cite{F-T,Olv,Ma,G-D,F-F,Fuc,Z-K,Fo}.
It is therefore natural to ask whether the same or similar
structures exist for the system of vortex filament.
In this paper we focus on structures that are described in
the language of differential geometry; we will investigate the
hereditary property \cite{F-F,Z-K} of the recursion operator,
Hamiltonian pair \cite{Ma,G-D}, and master symmetries \cite{Fuc}.
For these, the answers are all affirmative; the space $\BAL$ admits
these structures. The asymptotic boundary condition defining
$\BAL$ is critical for this result; a different situation is
encountered when the curve of a vortex filament is supposed to
be a loop \cite{Sa}.

The paper is organized as follows:
In Section~\ref{sect-2}, the definition of $\BAL$ is clarified.
It involves introducing a further condition to the conditions
defining $\BAL$ of \cite{L-P}.
Also in this section, some basic notions are described, and several
useful formulae for the variational calculus on $\BAL$ are summarized.
In Section~\ref{sect-3}, carefully specifing what are
admissible vector fields and what are admissible functionals,
we define a differential calculus on $\BAL$.
The calculus provides the framework for the subsequent analysis.
Section~\ref{sect-4}, consists of two subsections.
In Subsection~\ref{sect-4.1}, a recursion operator is shown to be
hereditary, and its several consequences are presented.
In Subsection~\ref{sect-4.2}, it is shown with the help of the
hereditary recursion operator that certain two operators form
a Hamiltonian pair.
In Section~\ref{sect-5}, master symmetries are investigated
and are applied to deriving an expression of constants
of motion in involution. This proves the inspection of Langer and
Parline.

\section{Balanced Asymptotically Linear Curves}
\label{sect-2}

Let $\APC$ be the space of infinitely extended, arclength-parametrized
smooth curves in the Euclidean space $\real^3$ with the standard
metric $\metric{\;}{}$.
We imply by the letter $\gamma$ an element of $\APC$ and
by $s$ the parameter for it; $s\mapsto\gamma(s)$ is a smooth
map $\real\to\real^3$ such that $\partial\gamma(s)/\partial s$
is a unit vector in the tangent space $T_{\gamma(s)}\real^3$.

A map $\APC\times\real\to\real$ is referred to as a scalar field
on $\APC$. A map $\XXX\colon\APC\times\real\to\coprod T_{\gamma(s)}
\real^3$ such that $\XXX(\gamma,\:s)\in T_{\gamma(s)}\real^3$ is
referred to as a tangent field (the term `vector field' is reserved
for the differential calculus). Here, $\coprod$ stands for the
direct-sum with respect to the index $(\gamma,\:s)\in\APC\times\real$.
Similar terminology is used also for a subset $\BAL$,
a space that we wish to manifest in this section.

The differential operator with respect to $s$ is denoted by $\ders$
(when acting on scalar fields) or by $\Ders$ (when acting on
tangent fields). These operators satisfy
\begin{eqnarray}
& \ders (fg) = (\ders f)g + f\ders g, & \label{eq.2}\\
& \Ders (f\XXX) = (\ders f) \XXX + f\Ders\XXX, \quad
 \ders\metric{\XXX}{\YYY} = \metric{\Ders\XXX}{\YYY}
  + \metric{\XXX}{\Ders\YYY} &
\end{eqnarray}
for all scalar fields $f$, $g$ and tangent fields $\XXX$, $\YYY$.
As in the equations above, we will often surpress the
argument $(\gamma,\:s)$.

A scalar field $F$ is called a functional if $F$ is independent
of $s$, {\it i.e.}, $\ders F = 0$.

We say a scalar field $f$ is asymptotically polynomial-like
if there exists a polynomial $P(s)\in\real[s]$ such that
$f(\gamma,\:s)/P(s)\to 0$ in the limit $s\to\pm\infty$ for every
curve $\gamma$. We say a scalar field $f$ is rapidly-decreasing
if $f(\gamma,\:s)P(s)$ for every polynomial $P(s)\in\real[s]$
converges to zero in the limit $s\to\pm\infty$ for every curve $\gamma$.

Let $f$ be a scalar field.
The scalar field $\ints f$ (anti-differentiation of $f$) and
the functional $\int\! f$ (definite integration of $f$) are defined by
\begin{eqnarray}
 (\ints f)(\gamma,\:s) &=& \frac{1}{2}\,
  \Bigl(
   \int_{-\infty}^s\! f(\gamma,\:\tilde{s})\ed \tilde{s}
  -\int_s^{\infty}\! f(\gamma,\:\tilde{s})\ed \tilde{s}
  \Bigr), \\
 (\int\! f)(\gamma)
  &=& \int_{-\infty}^{\infty}\! f(\gamma,\:\tilde{s})\ed \tilde{s}
\end{eqnarray}
provided that the integrations in the equations above converge.
In emplying operators $\ints$ and $\int$ in the following sections,
we will ensure the convergence by introducing certain rules.
It is easy to see
\begin{eqnarray}
& \ders\ints f = \ints\ders f = f, \quad
  \ders\int\! f = \int\!\ders f = 0,   & \label{eq.6} \\
& \ints(F f) = F\ints f,\quad
 \int\!(F f) = F\int\! f, & \label{eq.7}
\end{eqnarray}
where $f$ is a rapidly-decreasing scalar field
and $F$ is a functional.

We denote by $\TTT$, $\NNN$, $\BBB$ the tangent fields forming the
Fr\'enet frame, namely, $\TTT(\gamma,\:s)$, $\NNN(\gamma,\:s)$ and
$\BBB(\gamma,\:s)$ are orthonormal vectors in $T_{\gamma(s)}\real^3$
satisfying $\TTT(\gamma,\:s) = \partial\gamma(s)/\partial s$ and
the Fr\'enet-Serret relation
\begin{equation}
 \Ders \TTT = \kappa \NNN,\quad
 \Ders \NNN = -\kappa \TTT + \tau \BBB,\quad
 \Ders \BBB = -\tau \NNN.  \label{eq.8}
\end{equation}
Here, $\kappa$ and $\tau$ are scalar fields characterized by
(\ref{eq.8}), namely, $\kappa(\gamma)$ and $\tau(\gamma)$
are the curvature and torsion, respectively, of the curve $\gamma$.
Every tangent field is uniquely written as a linear combination
of $\TTT$, $\NNN$, $\BBB$ with the coefficients in scalar fields.

The space $\BAL$ introduced in \cite{L-P} is a subset of $\APC$ such that
(a) the curvature $\kappa(\gamma)$ of $\gamma\in\BAL$ is non-vanishing,
(b) $\gamma\in\BAL$ is asymptotic to a fixed line, {\it e.g.}, to
$z$-axis, and (c) ambiguity in the parametrization is completely
eliminated with imposing balancing condition.

To describe the balancing condition, we need to fix a reference curve
$\gamma_0\in\APC$ (or a reference line, $z$-axis) fulfilling the
asymptotic condition as in (b). The condition (b) says the existence
of functionals
$\length_\pm\colon\BAL\to\real$ with which the asymptotic behaviour
of $\gamma\in\BAL$ in the region $s\to\pm\infty$ is written as
$\gamma(s \pm \length_\pm (\gamma))\to\gamma_0(s)$.
With these functionals, the balancing condition (c) for $\gamma\in\BAL$
can be witten as $\length_+ (\gamma) = \length_- (\gamma)$.
The functional $\length := \length_+ + \length_-$
referred to as the renormalized length (relative to $\gamma_0$)
is well-defined, though the curve $\gamma\in\BAL\subset\APC$ is
of infinite length.

We supplement the condition (b) with prescribing how $\gamma\in\BAL$
converges to the reference curve $\gamma_0$; we suppose
\begin{equation}
 \left\{
  \begin{array}{l}
   \mbox{$\kappa$ is rapidly-decreasing, and}\\
   \mbox{$\kappa^{-1}\ders{}^n\kappa$ and $\ders{}^n\tau$,
    $n=0,\:1,\:\ldots$ are all asymptotically polynomial-like,}
  \end{array}
 \right.
\end{equation}
where $\kappa^{-1} := 1/\kappa$.

In the next section we introduce a differential calculus on $\BAL$.
The action of vector fields on functionals in this calculus
is defined to reproduce the usual variational calculus. Here we make
a few remarks on the variational calculus and give several useful
formulae. For more detailed description, we refer to \cite{L-P}.

Let $\XXX$ be a tangent field written as
\begin{equation}
 \XXX = \ints(\kappa g)\TTT + g \NNN + h \BBB  \label{eq.10}
\end{equation}
with certain rapidly-decreasing scalar fields $g,\:h$.
Below, $\XXX(\gamma)$ is identified with a variational vector field
along $\gamma$. The restriction on $\XXX$ mentioned above is
to force the variation to keep the arclength-parametrization
and conditions (b), (c).

In this paper the variational differential operator associated
with $\XXX$ of the form (\ref{eq.10}) is denoted by $\vari{\XXX}$
(when acting on scalar fields) or by $\Vari{\XXX}$ (when acting on
tangent fields).
For calculating the former, the following formulae are useful:
\begin{eqnarray}
& \vari{\XXX}(fg) = (\vari{\XXX}f)g + f\vari{\XXX}g, & \label{eq.11}\\
& \vari{\XXX}\ders f = \ders \vari{\XXX} f, \quad
  \vari{\XXX}\ints f = \ints \vari{\XXX} f, \quad
  \vari{\XXX}\int\! f = \int\! \vari{\XXX} f, &\\
& \vari{\XXX}\kappa = \metric{\NNN}{\Ders\Ders\XXX}, &\\
& \vari{\XXX}\tau
 = \ders\metric{\kappa^{-1}\BBB}{\Ders\Ders\XXX}
 + \metric{\kappa\BBB}{\Ders\XXX}, &\\
& \vari{\XXX}s = 0, &\label{eq.15}\\
& \vari{\XXX}\length = \int\! \metric{-\kappa\NNN}{\XXX}, &
\end{eqnarray}
where $f$ and $g$ are scalar fields. By abuse of notation, we
often use the letter $s$, by which we mean the scalar field
$\hat s$ such that $\hat s(\gamma,\:s) = s$, as we have done
in (\ref{eq.15}).
The latter can be calculated by the following formulae:
\begin{eqnarray}
& \Vari{\XXX}(f\YYY) = (\vari{\XXX}f)\YYY + f\Vari{\XXX}\YYY, &
 \label{eq.17} \\
& \Vari{\XXX}\TTT = \metric{\NNN}{\Ders\XXX}\NNN
 + \metric{\BBB}{\Ders\XXX}\BBB, &\\
& \Vari{\XXX}\NNN = \metric{\kappa^{-1}\BBB}{\Ders\Ders\XXX}\BBB
 - \metric{\NNN}{\Ders\XXX}\TTT, &\\
& \Vari{\XXX}\BBB = - \metric{\BBB}{\Ders\XXX}\TTT
 - \metric{\kappa^{-1}\BBB}{\Ders\Ders\XXX}\NNN, & \label{eq.20}
\end{eqnarray}
where $f$ is a scalar field and $\YYY$ is a tangent field.
These satisfy
\begin{equation}
 \vari{\XXX}\metric{\YYY}{\ZZZ}
  = \metric{\Vari{\XXX}\YYY}{\ZZZ}
  + \metric{\YYY}{\Vari{\XXX}\ZZZ}.
\end{equation}

\section{Differential Calculus}
\label{sect-3}

Let $A_n$, $n\in\integer$, be the $\ders$-invariant space ({\it i.e.},
$\ders f \in A_n$ for all $f\in A_n$) of scalar fields $f$ such that
$\kappa^{-n}f$ is asymptotically polynomial-like.
The elements of $A_n$ with $n > 0$ are rapidly-decreasing.

We notice the following properties possessed by $A_n$:
\begin{itemize}
\item[a1.] $A_n$, $\forall n\in\integer$, is an $\real$-vector space,
\item[a2.] $A_n \subset A_{n-1}$ (as $\real$-vector spaces)
 for all $n\in\integer$,
\item[a3.] $A_{-\infty} := A_0 \cup A_{-1} \cup \cdots$ is
 a commutative associative $\real$-algebra with the unit $1$,
\item[a4.] $fg \in A_{i+j}$ if $f\in A_i$, $g\in A_j$ for all
 $i,\:j\in\integer$,
\item[a5.] $\ders$ is an $\real$-linear operator
 such that $\ders(A_n)\subset A_n$
 for all $n\in\integer$,
\item[a6.] $\ints$ and $\int$ are $\real$-linear operators
 such that $\ints f \in A_0$ and $\int\! f \in A_0$
 for all $f \in A_2$,
\item[a7.] $\kappa \in A_1$, $\kappa^{-1}\in A_{-1}$,
 and $\tau,\:s,\:\length,\:1 \in A_0$,
\item[b1.] $\ders$ is a derivation of $A_{-\infty}$, {\it i.e.},
 Eq.~(\ref{eq.2}) hold for all $f,\:g\in A_{-\infty}$,
\item[b2.] Eqs.~(\ref{eq.6}) and (\ref{eq.7})
 hold for all $f\in A_2$ and $F\in \mbox{\rm Ker}\,\ders$,
\item[b3.] $\int\!(f \ints g) = - \int\!(g \ints f)$
 for all $f,\:g\in A_2$,
\item[b4.] $\kappa \kappa^{-1} = \ders s = 1$, and $\ders\length = 0$.
\end{itemize}

Let us consider the objects $\exprn{n}$ that are fully
characterized by the rules a1--a7 above; regarding a1--a7 (in which
$A_n$ should be read as $\exprn{n}$) as the axioms for $\exprn{n}$,
we define $\exprn{n}$, $n\in\integer$, as a family of objects
generated by the symbols or indeterminates
$\{\kappa,\:\kappa^{-1},\:\tau,\:s,\:\length,\:1\}$
with the algebraic operations. Here and in the following paragraph,
by algebraic operations we mean addition, scaling by a real number,
multiplication, $\ders$, $\ints$ and $\int$.
It is the implication of a6 that $\ints$ and $\int$ cannot act
on $\exprn{n}$, $n < 2$. By definition, rules (such as b1--b4)
not following from a1--a7 are not available for $\exprn{n}$.

Let $g_1$, $\ldots$, $g_r$ be independent variables running over
$\exprn{n_1}$, $\ldots$, $\exprn{n_r}$, respectively.
We say $f(g_1,\:\ldots,\:g_r)$ is an $\exprn{n}$-valued
variable algebraically depending on $g_1$, $\ldots$, $g_r$
if $f(g_1,\:\ldots,\:g_r)$ is an expression written in terms of
$\{g_1,\:\ldots,\:g_r,\:\kappa,\:\kappa^{-1},\:\tau,\:s,\:\length,\:1\}$
with use of the algebraic operations
and if the rules a1--a7 supplemented with the condition
$g_i\in\exprn{n_i}$ conclude $f(g_1,\:\ldots,\:g_r)\in\exprn{n}$.

We denote by $\alge{n}$ the space of scalar fields on $\BAL$ having
an expression that belongs to $\exprn{n}$. It is easy to see that
$\alge{n}$ is a subset of $A_n$.
The statements~a1--a7 and~b1--b4 remain true even if every
$A_n$ is read as $\alge{n}$. Moreover, these together with
the positive-definiteness or at least nondegeneracy of the
bi-linear form (\ref{eq.35}) are all of the fundamental setting
we need in constructing the theory developed in this paper.

\begin{proposition}
\label{prop.1}
Let $\Phi$ be an $\real$-linear map $\alge{1}\to\alge{0}$
induced from an $\real$-linear map $\exprn{1}\to\int\!(\exprn{2})$
in the apparent way. If this map is written as
$\Phi(g) = \int\! f(g)$ with an $\exprn{2}$-valued variable $f(g)$
algebraically depending on $g\in\exprn{1}$, then there exists
$h\in\alge{1}$ with which one can write $\Phi(g) = \int\! gh$
$\forall g\in\alge{1}$ as an equation in $\alge{0}$.
\end{proposition}

\begin{proof}
We note the formulae
\begin{equation}
 \int\!(f \ders g) = - \int\!(g \ders f), \quad
 \int\!(f \ints g) = - \int\!(g \ints f), \quad
 \int\!(f \int\! g) = \int\!(g \int\! f),  \label{eq.22}
\end{equation}
each of which is vaild as an equation in $\alge{0}$
if the left-hand side is given as an $\exprn{0}$-valued variable
algebraically depending on $f$ and $g$.
{}From bi-$\real$-linearity of multiplication and $\real$-linearity
of $\ders$, $\ints$ and $\int$, we see the existence of an expression
$\Phi(g) = \sum_i \int f_i(g)$ with $f_i(g)$ being $\exprn{2}$-valued
variables algebraically depending on $g\in\exprn{1}$ such that
no additions are used in the expression of $f_i(g)$.
Further, it is possible to suppose $g$ appears in each expression
$\int f_i(g)$ only once because of the $\real$-linearity of $\Phi$.
For such expressions, it is apparent how to apply successively the
formulae (\ref{eq.22}) to $\int f_i(g)$ to rewrite it into
the form $\int g h_i$. This process is justified if one considers
the equations in $\alge{0}$, while consideration in $\exprn{0}$ is
useful in verifying that the expressions $\int(\cdots)$ appearing in
each step make sense as $\exprn{0}$-valued variables and eventually
in deducing $h_i\in\alge{1}$.
\end{proof}

Let $\tang{n}$, $n\in\integer$, be the $\real$-vector space of tangent
fields defined by $\tang{n} := \{f\TTT + g\NNN + h\BBB\:|\: f\in
\alge{n-1},\; g,\:h \in\alge{n}\}$. It is easy to see that $\tang{n}$
is $\Ders$-invariant, {\it i.e.}, $\Ders(\tang{n})\subset\tang{n}$.

We denote by $\surj$ the surjection associated with
the identification $f\TTT \sim 0$ in $\tang{n}$, namely,
putting $\NN = \surj(\NNN)$, $\BB = \surj(\BBB)$, we write
\begin{equation}
 \surj(f\TTT + g\NNN + h\BBB) = g\NN + h\BB
\end{equation}
for scalar fields $f$, $g$, $h$.
The vector spaces $\surj(\tang{n})$ are left $\alge0$-modules with
$f(g\NN + h\BB) = (fg)\NN + (fh)\BB$ $\forall f\in\alge0$,
$\forall g,\:h\in\alge{n}$.

Let $\vect := \surj(\tang{1}) = \{g\NN + h\BB\:|\: g,\:h \in\alge{1}\}$.
Each element of $\vect$ is referred to as a vector field on $\BAL$.
Through the injection $\injec\colon\vect\to\tang{1}$ defined by
\begin{equation}
 \injec(g\NN + h\BB)
  := \ints(\kappa g)\TTT + g\NNN + h\BBB,
\end{equation}
a vector field $\XX$ induces a derivation
--- variational differential associated with $\injec(\XX)$.
This derivation acting on $\alge{n}$ or $\tang{n}$ can be evaluated
with the formulae (\ref{eq.11})--(\ref{eq.20}).

\begin{proposition}
\label{prop.2}
The vector spaces $\alge{n}$ and $\tang{n}$ are invariant under
the action of the vector fields, namely,
$\vari{\injec(\XX)}(\alge{n})\subset\alge{n}$ and
$\Vari{\injec(\XX)}(\tang{n})\subset\tang{n}$
$\forall \XX\in\vect$, $\forall n\in\integer$.
\end{proposition}

\begin{proof}
It is essential that every vector field $\XX = g\NN + h\BB$ is
written with $g,\:h\in\alge{1}$. Taking notice of this situation,
we find that the formulae (\ref{eq.11})--(\ref{eq.20})
ensure the invariance of $\alge{n}$ and $\tang{n}$ under the
action of vector fields.
\end{proof}

The space $\vect$ of vector fields is a Lie algebra, and $\alge{n}$
are $\vect$-modules. This is an immediate consequence of the
following theorem.
\begin{theorem}
\label{th.3}
The $\real$-vector space $\injec(\vect)$ is a Lie algebra with
the product
\begin{equation}
 \cp{\XXX}{\YYY} := \Vari{\XXX}\YYY - \Vari{\YYY}\XXX \quad
  \forall \XXX,\:\YYY\in\injec(\vect).
\end{equation}
For every $n\in\integer$, the algebra $\alge{n}$ is
a $\injec(\vect)$-module with the action
$\vari{\XXX}$, $\XXX\in\injec(\vect)$, namely,
\begin{equation}
 (\vari{\XXX} \vari{\YYY} - \vari{\YYY} \vari{\XXX}
  - \vari{\cp{\XXX}{\YYY}}) f = 0 \quad
  \forall \XXX,\:\YYY\in\injec(\vect),\;
  \forall f\in\alge{n}.  \label{eq.26}
\end{equation}
\end{theorem}
\begin{proof}
The statements are verified by using
(\ref{eq.11})--(\ref{eq.20}). A convenient procedure
is as follows: First, verify that $\cp{\XXX}{\YYY}$ belongs
to $\injec(\vect)$ for all $\XXX,\:\YYY\in\injec(\vect)$. Second,
show the equation (\ref{eq.26}) for $f = \kappa$, $\tau$,
$s$, $\length$ and then extend (\ref{eq.26}) to the whole
$\alge{-\infty} = \alge{0}\cup\alge{-1}\cup\cdots$.
Finally, verify the Jacobi identity in $\injec(\vect)$ with the
help of (\ref{eq.26}).
\end{proof}
The theorem above is quite similar to Theorem~1 of \cite{Sa}
in particular in the proof,
though the considered objects are different.

To simplify expressions, we put
\begin{equation}
 \VARI{\XX} := \surj \circ \Vari{\injec(\XX)} \circ \injec
\end{equation}
for $\forall\XX\in\vect$, so that we can write the commutator
product of $\vect$ as
\begin{equation}
 \cp{\XX}{\YY} = \VARI{\XX}\YY - \VARI{\YY}\XX \quad
 \forall \XX,\:\YY\in\vect.
\end{equation}
Likewise we put
\begin{equation}
 \inner{g_1\NN + h_1\BB}{g_2\NN + h_2\BB} := g_1 g_2 + h_1 h_2,
  \label{eq.29}
\end{equation}
which defines a bi-linear form
$\surj(\tang{i})\times\surj(\tang{j})\to\alge{i+j}$.
Then, we have
\begin{eqnarray}
 \VARI{\XX}\YY &=& (\vari{\injec(\XX)} \inner{\NN}{\YY})\NN
  + (\vari{\injec(\XX)} \inner{\BB}{\YY})\BB
  + (\ints\inner{\kappa\NN}{\YY}) \surj\Ders\injec(\XX) \nonumber \\
&&
 {} - \inner{\kappa^{-1}\BB}{\surj\Ders\Ders\injec(\XX)}
  (\inner{\BB}{\YY} \NN - \inner{\NN}{\YY}\BB)
 \label{eq.30}
\end{eqnarray}
for all $\XX,\:\YY\in\vect$.

Let $\func$ be the subalgebra of $\alge0$ generated by $1$, $\length$
and the elements of $\int\!(\alge2)$. The vector space $\func$ is
an $\vect$-submodule of $\alge0$, {\it i.e.}, $\vari{\injec(\XX)}(\func)
\subset\func$ $\forall\XX\in\vect$.
We denote the action of $\vect$ on $\func$ by the left-action,
namely, $\XX F = \vari{\injec(\XX)} F$ $\forall \XX\in\vect$,
$\forall F\in\func$. This action is a derivation:
\begin{equation}
 \XX(FG) = (\XX F)G + F(\XX G)
  \quad\forall \XX\in\vect,\; \forall F,\:G\in\func. \label{eq.31}
\end{equation}
Since $\func\subset\alge0$, we see that $\vect$ is a left $\func$-module.
Taking notice of $\ders F = 0$ $\forall F\in\func$ and referring
(\ref{eq.11})--(\ref{eq.20}), we easily find
$\vari{\injec(F\XX)}g = F \vari{\injec(\XX)}g$,
$\VARI{F\XX}\YY = F \VARI{\XX}\YY$ and
$\VARI{\XX}(F\YY) = (\vari{\injec(\XX)}F)\YY + F\VARI{\XX}\YY$
$\forall F\in\func,\;\forall\XX,\:\YY\in\vect,\;\forall g\in\alge{n}$.
Hence we see
\begin{eqnarray}
& (F \XX) G = F (\XX G) \quad \forall F,\:G\in\func,\;
 \forall\XX\in\vect, & \\
& \Ld{\XX}(F \YY) = (\Ld{\XX}F)\YY + F \Ld{\XX}\YY \quad
 \forall\XX,\:\YY\in\vect,\; \forall F\in\func, &  \label{eq.33}
\end{eqnarray}
where
\begin{equation}
 \Ld{\XX}\YY := \cp{\XX}{\YY}, \quad
 \Ld{\XX} F := \XX F \quad
 \forall \XX,\:\YY\in\vect,\; \forall F\in\func.
   \label{eq.34}
\end{equation}

Below, we construct in the usual, algebraic manner a differential
calculus, in which the algebra $\func$ consisting of functionals
on $\BAL$ plays the role of the algebra of functions.
The construction is based on the pair
$(\func,\:\vect)$ of commutative algebra and Lie algebra.
It is essential for this construction that $\func$ is a left
$\vect$-module, $\vect$ is a left $\func$-module, and
the equations (\ref{eq.31})--(\ref{eq.33}) hold.
We would like to make a further remark.
Let $\Riem\colon\vect\times\vect\to\func$ be a symmetric form
defined by
\begin{equation}
 \Riemann{\XX}{\YY} := \int\, \inner{\XX}{\YY}
   \label{eq.35}
\end{equation}
with (\ref{eq.29}). This is bi-$\func$-linear and
positive-definite. We refer to $\Riem$ as the Riemannian structure
on $\BAL$. Given $F\in\func$, the vector field $\XX\in\vect$ such that
$\YY F = \Riemann{\XX}{\YY}$ $\forall \YY\in\vect$ is called the
gradient of $F$ and is denoted by $\grad F$. The existence of the
gradient for every element of $\func$ can be verified by virtue
of Propositions~\ref{prop.1} and~\ref{prop.2}.
In contrast to the differential calculus on finite dimensional
Riemannian manifolds, this seems to be quite nontrivial.
This situation is necessary for realizing the space of 1-forms
as a space identifiable with $\vect$.

Let $\form{p}$ denote the vector space of maps $\eta\colon\vect^{
\times p}\to\func$ such that $F := \eta(\UU_1,\:\cdots,\:\UU_p)$ with
$\UU_i = g_i\NN + h_i\BB\in\vect$ is $\func$-linear in each $\UU_i$,
skew-symmetric (if $p\geq2$) under the exchange of $\UU_i$ and $\UU_j$,
$i\neq j$, and $F$ can be expressed as an $\exprn{0}$-valued variable
algebraically depending on $g_1,\:h_1,\:\ldots,\:g_p,\:h_p$.
Such a map $\eta\in\form{p}$ is referred to as a $p$-th order
differential form or $p$-form on $\BAL$.
{}From Proposition~\ref{prop.1} and the nondegeneracy
of (\ref{eq.35}), we see that for every 1-form $\xi$
there uniquely exists a vector field $\XX$ such that
$\xi(\YY) = \Riemann{\XX}{\YY}$ $\forall \YY\in\vect$.

The exterior derivative is a map
$\ed\colon\form{p}\rightarrow\form{p+1}$
defined by
\begin{eqnarray}
 (\ed \eta)(\UU_0,\:\ldots,\:\UU_p) &=&
   \sum_{i=0}^{p} (-1)^{i} \UU_i(\eta(\UU_0,\:\ldots,\:
   \check{\UU_i},\:\ldots,\:\UU_p)) \nonumber \\
   &+& \sum_{i<j} (-1)^{i+j}\eta(\cp{\UU_i}{\UU_j},\:
   \UU_0,\:\ldots,\:\check{\UU_i},\:\ldots,\:\check{\UU_j},\:
   \ldots,\:\UU_p)
   \label{eq.36} \\
 && \qquad\qquad\qquad\qquad\qquad\qquad\qquad
  \forall \eta\in\form{p},\;\forall \UU_i\in\vect, \nonumber
\end{eqnarray}
where $\check{\UU_i}$ stands for the absense of $\UU_i$.
This is a coboundary operator, {\it i.e.}, $\ed\circ\ed = 0$.
The interior product $\ip{\XX}\colon\form{p+1}\to\form{p}$ with
$\XX\in\vect$ is defined by
\begin{equation}
 (\ip{\XX} \eta)(\UU_1,\:\ldots,\:\UU_p)
  = \eta(\XX,\:\UU_1,\:\ldots,\UU_p).
\end{equation}
The Lie derivative $\Ld{\XX}$ in the direction of $\XX\in\vect$
is an opeator acting on each $\vect$-module consisting of certain
$\func$-vectors, so-called tensor fields, and $\XX\mapsto \Ld{\XX}$
provides a representation of the Lie algebra $\vect$.
The definition was already given both on $\func$ and $\vect$ in
(\ref{eq.34}). The extension to other tensor fields
is made by imposing Leibnitz rule. For example, if
$R$ is an $\func$-linear map $\vect\to\vect$, then
$\Ld{\XX}(R\YY) = (\Ld{\XX}R)\YY + R\Ld{\XX}\YY$
$\forall \XX,\:\YY\in\vect$. For a $p$-form $\eta$, the formula
\begin{equation}
  \Ld{\XX}\eta = \ip{\XX} \ed \eta + \ed \ip{\XX} \eta
   \label{eq.38}
\end{equation}
is available. It is possible to introduce the exterior product,
which is, however, not used in this paper.

\section{Recursion Operator and Hamiltonian Pair}
\label{sect-4}

An $\func$-linear map $K\colon\vect\to\vect$ is called a skew-adjoint
operator if $\Riemann{K\XX}{\YY} = - \Riemann{\XX}{K\YY}$.
With a skew-adjoint operator $J$ defined by
\begin{equation}
  J(\XX) = \inner{\BB}{\XX} \NN - \inner{\NN}{\XX} \BB,
  \label{eq.39}
\end{equation}
the Marsden-Weinstein Poisson structure \cite{M-W} can be written as
$\Poisson{F}{G} = (J\grad F)G$ $\forall F,\:G\in\func$.
The operator $J$, $J^2 = -1$, is a complex structure;
it can be shown by a direct calculation that the Nijenhuis torsion
(see Eq.~(\ref{eq.43})) of $J$ vanishes, though this fact
is not used in this paper.
The vortex filament equation (\ref{eq.1}) can be understood
as a Hamiltonian equation with the Hamiltonian functional $\length$;
indeed
\begin{equation}
 \kappa\BBB = \injec(J \grad \length).  \label{eq.40}
\end{equation}

Making use of the Hasimoto map, Langer and Perline \cite{L-P} found
that the vector fields $\kappa\BB$, $KJ^{-1}(\kappa\BB)$,
$(KJ^{-1})^2(\kappa\BB)$, $\ldots$ are Hamiltonian flows
(see Subsection~\ref{sect-4.2}) associated with the constants of motion
in involution, where $K$ is another skew-adjoint operator defined by
\begin{equation}
 K(\XX) = J \surj \Ders \injec J (\XX).
  \label{eq.41}
\end{equation}
The operator
\begin{equation}
 R := K J^{-1} = J \circ \surj \circ \Ders \circ \injec
  \label{eq.42}
\end{equation}
is referred to as the recursion operator.
It should be emphasized that the definition of $\vect$ given in the
preceding section is consistent with $J$, $K$ and $R$, namely,
these opertors make sense as $\func$-linear maps $\vect\to\vect$.

Below, after giving several results on the recursion operator $R$
(Subsection~\ref{sect-4.1}), we shall show that $J$ and $K$ form
a Hamiltonian pair (Subsection~\ref{sect-4.2}).
The approach pursued in this section is similar
to that of \cite{Z-K}.

\subsection{The hereditary property}
\label{sect-4.1}

Let $R$ be an $\func$-linear map $\vect\to\vect$.
In most statements of this subsection, we need not suppose $R$ is
the recursion operator defined by (\ref{eq.42});
the exception is Theorem~\ref{th.4}.

The Nijenhuis torsion $N_R$ of an $\func$-linear operator
$R\colon\vect\to\vect$ is an $\func$-linear map $\vect\times\vect\to\vect$
defined by
\begin{eqnarray}
 N_R(\XX,\:\YY) & := & (\Ld{R\XX} R - R \Ld{\XX} R) \YY \nonumber \\
 & = & \cp{R\XX}{R\YY} - R \cp{R\XX}{\YY} - R \cp{\XX}{R\YY}
   + R^2 \cp{\XX}{\YY}.  \label{eq.43}
\end{eqnarray}

An $\func$-linear operator $R$ such that
$N_R = 0$ is said to be hereditary \cite{Olv,F-F}; see also
\cite{Ma,G-D,Z-K}.
For a hereditary operator $R$, it is easy to see that
$\Ld{R\XX}(R^n) = R\Ld{\XX}(R^n)$ and further
$\Ld{R^m \XX}(R^n) = R^m \Ld{\XX}(R^n)$, namely,
\begin{equation}
 \cp{R^m \XX}{R^n \YY} - R^n \cp{R^m \XX}{\YY}
 = R^m \cp{\XX}{R^n \YY} - R^{m+n} \cp{\XX}{\YY}
 \label{eq.44}
\end{equation}
for all vector fields $\XX$, $\YY$.

\begin{theorem}
\label{th.4}
The recursion operator $R$ defined by (\ref{eq.42}) is hereditary.
\end{theorem}

\begin{proof}
For all $\XX\in\vect$, define $\VARI{\XX}R \colon\vect\to\vect$ by
\begin{equation}
 (\VARI{\XX}R)(\YY) = \VARI{\XX}(R \YY) - R(\VARI{\XX}\YY)
 \quad \forall \YY\in\vect,
\end{equation}
which is found to be
\begin{equation}
 (\VARI{\XX}R)(\YY) = (\ints\inner{\kappa\NN}{R\YY}) J^{-1} R \XX
  + (\ints\inner{R\XX}{R\YY}) \kappa\BB  \label{eq.46}
\end{equation}
by using (\ref{eq.30}).
Substituting the formula above into
\[
 N_R(\XX,\:\YY) = (\VARI{R\XX}R)(\YY) - R((\VARI{\XX}R)(\YY)
  - (\XX \leftrightarrow \YY),
\]
we find $N_R = 0$.
\end{proof}

Let $\eta$ be a p-form, $p\geq 1$. We say $\eta$ is compatible
with $R$ if
\begin{equation}
 \ip{R\YY}\ip{\XX}\eta = \ip{\YY}\ip{R\XX}\eta
  \quad \forall \XX,\:\YY\in\vect.
\end{equation}
By definition, every 1-form is compatible with $R$.
Suppose $\eta$ is a $p$-form compatible with $R$.
Then, it is possible to define a $p$-form $\eta\circ R$ such that
$\ip{\XX}(\eta\circ R) = \ip{R \XX}\eta$ for all $\XX\in\vect$.
Further, $\eta\circ R$ is again compatible with $R$.
Hence a $p$-form compatible with $R$ induces $p$-forms
$\eta\circ R^n$ compatible with $R$, $n=0,\:1,\:2,\:\ldots$,
such that
\begin{equation}
 \ip{\XX}(\eta\circ R^n) = \ip{R^n \XX}\eta \quad \forall \XX\in\vect.
\end{equation}

For a $p$-form $\eta$ compatible with $R$, we have the formulae
\begin{eqnarray}
& \Ld{R^n \XX}\eta = \ip{R^n \XX} \ed\eta
 + \Ld{\XX}(\eta\circ R^{n}) - \ip{\XX} \ed(\eta\circ R^{n}), &
  \label{eq.49} \\
& \ip{\YY}\Ld{\XX} (\eta\circ R) = \ip{R\YY}\Ld{\XX} \eta
  + \ip{(\Ld{\XX}R)\YY} \eta, &
  \label{eq.50}
\end{eqnarray}
where $n$ is an arbitrary non-negative integer and $\XX$ is an
arbitrary vector field.
The former (\ref{eq.49}) is an immedate consequense of
(\ref{eq.38}).
The latter (\ref{eq.50}) is nothing but the Leibnitz rule
\[
  (\Ld{\XX} (\eta\circ R)) (\YY,\:\UU_2,\:\ldots,\:\UU_p)
  = (\Ld{\XX} \eta) (R\YY,\:\UU_2,\:\ldots,\:\UU_p)
  + \eta((\Ld{\XX}R)\YY,\:\UU_2,\:\ldots,\:\UU_p).
\]
For a 2-form $\omega$ compatible with $R$, we have
\begin{eqnarray}
&& \big\{ (\Ld{R^n \XX}\omega)(\YY,\:\ZZ)
 - (\omega\circ  R^n)(\XX,\:\cp{\YY}{\ZZ}) \big\}
 + \cycle(\XX,\:\YY,\:\ZZ) \nonumber \\
&=& \big\{ \ed\omega(R^n\XX,\:\YY,\:\ZZ)
 + \cycle(\XX,\:\YY,\:\ZZ) \big\}
 - 2 \ed(\omega\circ R^n)(\XX,\:\YY,\:\ZZ),
 \label{eq.51}
\end{eqnarray}
where $n$ is an arbitrary non-negative integer and
$\XX$, $\YY$, $\ZZ$ are arbitrary vector fields.
The formula (\ref{eq.51}) is verified as follows:
Starting with the substitution of (\ref{eq.49}), we calculate
the left-hand side of (\ref{eq.51}) as
\begin{eqnarray*}
&& \ed\omega(R^n \XX,\:\YY,\:\ZZ)
 + (\Ld{\XX}(\omega\circ R^{n}))(\YY,\:\ZZ)
 - \ed(\omega\circ R^{n}) (\XX,\:\YY,\:\ZZ) \\
&& {}- (\omega\circ R^n)(\XX,\:\cp{\YY}{\ZZ})
 + \cycle(\XX,\:\YY,\:\ZZ) \\
&=& \big\{ \ed\omega(R^n \XX,\:\YY,\:\ZZ)
 + (\Ld{\XX}(\omega\circ R^{n}))(\YY,\:\ZZ)
 - (\omega\circ R^n)(\YY,\:\cp{\ZZ}{\XX}) \\
&& {}+ \cycle(\XX,\:\YY,\:\ZZ) \big\}
 - 3 \ed(\omega\circ R^{n}) (\XX,\:\YY,\:\ZZ) \\
&=&  \big\{ \ed\omega(R^n \XX,\:\YY,\:\ZZ)
 + \XX((\omega\circ R^{n}))(\YY,\:\ZZ))
 - (\omega\circ R^n)(\cp{\XX}{\YY},\:\ZZ) \\
&& {}+ \cycle(\XX,\:\YY,\:\ZZ) \big\}
 - 3 \ed(\omega\circ R^{n}) (\XX,\:\YY,\:\ZZ).
\end{eqnarray*}
Then, recalling the definition (\ref{eq.36}) of exterior
derivative, we see this
equals to the right-hand side of (\ref{eq.51}).

\begin{lemma}
\label{lemm.5}
Let $R$ be an $\func$-linear map $\vect\to\vect$. If $\eta$ is
a $p$-form compatible with $R$, then
\begin{equation}
 \ip{\YY} \ip{\XX} \ed (\eta\circ R^2) 
 - \ip{R\YY} \ip{\XX} \ed (\eta\circ R)
 - \ip{\YY} \ip{R\XX} \ed (\eta\circ R)
 + \ip{R\YY} \ip{R\XX} \ed \eta
 = - \ip{N_R(\XX,\:\YY)} \eta    \label{eq.52}
\end{equation}
for all vector fields $\XX,\:\YY$.
\end{lemma}

\begin{proof}
That the left-hand side of (\ref{eq.52}) equals to
\[
  \ip{\YY} \Ld{\XX} (\eta\circ R^2) - \ip{R\YY} \Ld{\XX} (\eta\circ R)
 - \ip{\YY} \Ld{R\XX} (\eta\circ R) + \ip{R\YY} \Ld{R\XX} \eta
\]
can be shown by the substitution of the identity $\ip{\YY} \ip{\XX} \ed
 = \ip{\YY} \Ld{\XX} - \Ld{\YY} \ip{\XX} + \ed \ip{\YY} \ip{\XX}$
$\forall \XX,\:\YY\in\vect$ following from (\ref{eq.38}).
Using (\ref{eq.50}), we can rewrite the expression above
into $\ip{(\Ld{\XX}R)\YY} (\eta\circ R) - \ip{(\Ld{R\XX}R)\YY} \eta$,
which obviously equals to the right-hand side of (\ref{eq.52}).
\end{proof}

\subsection{Schouten bracket and Hamiltonian pair}
\label{sect-4.2}

Let us recall the notion of a Hamiltonian operator/pair \cite{Ma,G-D}.

Let $H_i$, $i=0,\:1$ be $\func$-linear maps $\form{1}\to\vect$.
Suppose $H_i$ are skew-symmetric, {\it i.e.},
\begin{equation}
 \xi(H_i\eta) = - \eta(H_i\xi) \quad \forall \xi,\:\eta\in\form{1}.
 \label{eq.53}
\end{equation}
The Schouten bracket $\Sb{H_0}{H_1}$ between $H_0$ and $H_1$ is
a skew-symmetric tri-$\func$-linear
map $\form{1}\times\form{1}\times\form{1}\to\func$ defined by
\begin{equation}
 \Sb{H_0}{H_1}(\xi,\;\eta,\:\zeta)
  = \xi(H_0 \Ld{H_1\eta} \zeta) + (H_0 \leftrightarrow H_1)
   + \cycle(\xi,\;\eta,\:\zeta).
\end{equation}

A skew-symmetric $\func$-linear map $H_0\colon\form{1}\to\vect$ is
referred to as a Hamiltonian operator if $\Sb{H_0}{H_0} = 0$.
The vector field $H_0\ed F$ associated with $F\in\func$
via Hamiltonian operator $H_0$ is called the Hamiltonian vector
field of $F$.
A Hamiltonian operator $H_0$ induces the Poisson structure
\begin{equation}
 \Poisson{F}{G} {}_{H_0} = (H_0\ed F)G \quad\forall F,\:G\in\func
  \label{eq.55}
\end{equation}
and $H_0 \circ \ed$ is a morphism $\func\to\vect$ of Lie algebras.
Hamiltonian operators $H_0$ and $H_1$ are said to be
a Hamiltonian pair if $\Sb{H_0}{H_1} = 0$.

The existence of a Hamiltonian pair with certain conditions implies
the integrability --- the existence of a sequence of functionals in
involution or Poisson-commutative functionals.

Returning to the case of $\BAL$, we define
$H_n\colon\form{1}\to\vect$, $n = 0,\:1,\:\ldots$ by
\begin{equation}
 H_n = R^n H_0, \quad
 \Riemann{\XX}{H_0 \eta} = - \eta (J \XX)
 \quad \forall \XX\in\vect,\;\forall \eta\in\form{1}
 \label{eq.56}
\end{equation}
with $J$ and $R$ defined by (\ref{eq.39}) and (\ref{eq.42}).
Under the identification caused by the Riemannian structure $\Riem$,
we see that $H_0$ and $H_1$
are nothing but the operators $J$ and $K$, respectively.
Indeed, $H_n\circ\ed = R^n J\circ\grad$.

We shall show that $H_m$ and $H_n$ form a Hamiltonian pair.
For this aim, it is useful to introduce the sequence of 2-forms
$\Omega_n$, $n =0,\:1,\:\ldots$,
\begin{equation}
 \Omega_n = \Omega_0 \circ R^n, \quad
 \Omega_0(\XX,\:\YY) = \Riemann{J^{-1}\XX}{\YY}
 \quad \forall \XX,\:\YY\in\vect.  \label{eq.57}
\end{equation}
The well-definedness of $\Omega_n$ as 2-forms is explained as follows:
As was mentioned, $J$ and $K$ are skew-adjoint operators.
{}From the skew-adjointness of $J$, we see that $\Omega_0$ is
well-defined as a 2-form. From the skew-adjointness of $J$ and $K$,
we see that $\Omega_0$ is compatible with $R$.
Hence, $\Omega_n = \Omega_0\circ R^n$ are well-defined as 2-forms.
These 2-forms are related to $H_n$ in the following way:
\begin{equation}
 \Omega_m(H_n \xi,\: \YY) = \xi(R^{m+n} \YY)
 \quad \forall \xi\in\form{1},\; \forall \YY\in\vect.
  \label{eq.58}
\end{equation}

It is possible to show that $\ed\Omega_0 = \ed\Omega_1 = 0$ by
using (\ref{eq.30}).
Since $R$ is hereditary, we find
\begin{equation}
 \ed\Omega_0 = \ed\Omega_1 = \ed\Omega_2 = \cdots = 0
  \label{eq.59}
\end{equation}
as a consequence of Lemma~\ref{lemm.5}.
We note that $\ed\Omega_0 = 0$ (and Theorem~\ref{th.6} for the
case $m = n = 0$) is implied in \cite{M-W}, because
$\Omega_0$ is the symplectic structure corresponding
to the Hamiltonian operator $J$.
It should be mentioned that $\Omega_n$, $n \geq 1$, is not
the symplectic structure corrsponding to $H_n$.

\begin{theorem}
\label{th.6}
Two operators arbitrarily chosen from the sequence $H_n$ defined by
(\ref{eq.56}) form a Hamiltonian pair, {\it i.e.},
$\Sb{H_m}{H_n} = 0$ for all non-negative integers $m$ and $n$.
\end{theorem}
This theorem follows immediately from (\ref{eq.59})
and the lemma below.

\begin{lemma}
\label{th.7}
Let $R\colon\vect\to\vect$ be a hereditary operator
and $\Omega_0$ a 2-form compatible with $R$.
Suppose the map $\vect\to\form{1}$, $\XX\mapsto\ip{\XX}\Omega_0$
is invertible, so that an $\func$-linear map $H_0\colon \form{1}\to\vect$
is defined by $\Omega_0(H_0\xi,\:\YY) = \xi(\YY)$
$\forall\xi\in\form{1},\;\forall\YY\in\vect$.
Then, $H_n := R^n H_0$, $n = 0,\:1,\:2,\:\ldots$
are skew-symmetric $\func$-linear maps $\form{1}\to\vect$
and the Schouten brackets between them are written as
\begin{eqnarray}
&& \Sb{H_m}{H_n}(\xi,\:\eta,\:\zeta) \nonumber \\
&=& 4 \ed\Omega_{m+n}(H_0\xi,\:H_0\eta,\:H_0\zeta) \label{eq.60} \\
&& {}- \big\{ \ed\Omega_{m}(H_n\xi,\:H_0\eta,\:H_0\zeta)
   + (m\leftrightarrow n) + \cycle(\xi,\:\eta,\:\zeta) \big\}
 \nonumber
\end{eqnarray}
with $\Omega_n := \Omega_0\circ R^n$.
\end{lemma}

\begin{proof}
The $\func$-linearity of $H_n$ is apparent. Further, the
calculation
\begin{eqnarray*}
 \xi(H_n\eta) &=& \Omega_0(H_0\xi,\:H_n\eta)
 = (\Omega_0\circ R^n)(H_0\xi,\:H_0\eta) \\
 &=& - (\Omega_0\circ R^n)(H_0\eta,\:H_0\xi)
 = - \Omega_0(H_0\eta,\:H_n\xi) = - \eta(H_n\xi)
\end{eqnarray*}
shows that $H_n$ are skew-symmetric. The Schouten bracket
$\Sb{H_m}{H_n}$ is therefore well-defined and is calculated
as follows: First, we notice
\[
 \xi(H_m \Ld{H_n\eta} \zeta)
 = - (\Ld{H_m\eta} \zeta)(H_n\xi)
 = - (H_m\eta)(\zeta(H_n\xi))
     + \zeta(\cp{H_m\eta}{H_n\xi}),
\]
so that we have
\[
 \Sb{H_m}{H_n}(\xi,\;\eta,\:\zeta)
 = \big\{ - (H_m\eta)(\zeta(H_n\xi)) + \zeta(\cp{H_m\eta}{H_n\xi})
 \big\}  + (m \leftrightarrow n) + \cycle(\xi,\;\eta,\:\zeta).
\]
Using (\ref{eq.44}), we see
\begin{eqnarray*}
 \Sb{H_m}{H_n}(\xi,\;\eta,\:\zeta)
&=& \big\{ - (H_m\eta)(\Omega_n(H_0\zeta,\:H_0\xi))
  + \Omega_n(H_0\zeta,\:\cp{H_m\eta}{H_0\xi}) \\
&& {}+ \Omega_m(H_0\zeta,\:\cp{H_0\eta}{H_n\xi})
  - \Omega_{m+n}(H_0\zeta,\:\cp{H_0\eta}{H_0\xi}) \big\} \\
&& {}+ (m \leftrightarrow n) + \cycle(\xi,\;\eta,\:\zeta).
\end{eqnarray*}
Taking notice of the symmetry under the exchange $m\leftrightarrow n$
and the cyclic permutation, we see that the first three terms in
the right-hand side sum up
to $(- \Ld{H_m\xi}\Omega_n)(H_0\eta,\:H_0\zeta)$.
Then, with the help of (\ref{eq.51}) we obtain (\ref{eq.60}).
\end{proof}

\section{Symmetries and Master Symmetries}
\label{sect-5}

Let $\XX_n$, $n = 0,\:1,\:2,\:\ldots$ be the vector fields
\begin{equation}
 \XX_n = R^n(\kappa \BB)  \label{eq.61}
\end{equation}
and $\YY_n$, $n = 1,\:2,\:3,\:\ldots$ the vector fields
\begin{equation}
 \YY_n = R^{n-1}(s \kappa \BB),  \label{eq.62}
\end{equation}
where $R$ is the recursion operator defined by (\ref{eq.42}).
The vector fields $\XX_n$ are those given in \cite{L-P} with
a difference in their index (shifted by 2).

\begin{lemma}
The vector fields $\XX_0$ and $\YY_1$ act on the recursion operator
$R$ of (\ref{eq.42}) as follows:
\begin{equation}
 \Ld{\XX_0}R = 0,\quad \Ld{\YY_1}R = - R^2.
\end{equation}
\end{lemma}
\begin{proof}
This is shown through a somewhat tedious calculation.
It is easy to see
$(\Ld{\XX}R)(\YY) = (\VARI{\XX}R)(\YY) - \VARI{R\YY}\XX + R\VARI{\YY}\XX$
for all vector fields $\XX$, $\YY$. We continue the calculation
with substitution of (\ref{eq.30}) and (\ref{eq.46}), and
finally arrive at the lemma.
\end{proof}

As a corollary of the lemma, it is immediate to see
\begin{equation}
 \Ld{\XX_0}R^n = 0, \quad
 \Ld{\YY_1}R^n = - n R^{n+1}. \label{eq.64}
\end{equation}

\begin{proposition}
The vector fields $\XX_0,\:\XX_1,\:\ldots$ and $\YY_1,\:\YY_2,\:\ldots$
form a Lie subalgebra of $\vect$ such that
\begin{eqnarray}
 \cp{\XX_n}{\XX_m} &=& 0,  \label{eq.65} \\
 \cp{\XX_n}{\YY_m} &=& (n+2)\XX_{n+m},  \label{eq.66} \\
 \cp{\YY_n}{\YY_m} &=& (n-m)\YY_{n+m}.  \label{eq.67}
\end{eqnarray}
\end{proposition}

\begin{proof}
As was stated in Theorem~\ref{th.4} the recursion
operator $R$ is hereditary, hence the formula (\ref{eq.44})
is available. This formula can be written also in the following
form:
\[
 \cp{R^m \XX}{R^n \YY} = - R^n (\Ld{\YY} R^m) \XX
 + R^m (\Ld{\XX} R^n) \YY + R^{m+n} \cp{\XX}{\YY}
 \quad \forall \XX,\:\YY\in\vect.
\]
Substituing (\ref{eq.64}) to the formula above, we
obtain (\ref{eq.65}) and (\ref{eq.67}).
We can show (\ref{eq.66}) in much the same way with the help of
\begin{equation}
 \cp{\XX_0}{\YY_1} = 2 \XX_1.
\end{equation}
This equation can be verified by using (\ref{eq.30}).
\end{proof}

It is immediate from (\ref{eq.1}) to see that $\XX_0$ is
the flow of the vortex filament equation.
A vector field commuting with the flow $\XX_0$
of an evolution equation is,
generally, called a symmetry (of the equation).
Thus $\XX_n$ of (\ref{eq.61}) can be described as symmetries of
the vortex filament equation.
Since these symmetries are generated from $\XX_0$ by the action of
$\YY_n$ as in (\ref{eq.66}), vector fields $\YY_n$ are
referred to as master symmetries \cite{Fuc}.
Sometimes the term `recursion operator' is used for meaning
an $\func$-linear operator $R\colon\vect\to\vect$ such that
$\Ld{\XX_0}R = 0$, {\it i.e.}, an operator
that maps a symmetry into another symmetry \cite{Olv}.
The operator $R$ of (\ref{eq.42}) is a recursion operator also
in this sense.

\begin{proposition}
By means of the Lie derivative, symmetries (\ref{eq.61}) and
master symmetries (\ref{eq.62}) act on the 2-forms $\Omega_n$
defined by (\ref{eq.57}) as follows:
\begin{eqnarray}
 \Ld{\XX_{m-1}}\Omega_n &=& 0, \label{eq.69} \\
 \Ld{\YY_m}\Omega_n &=& (3 - m - n) \Omega_{m+n},
  \label{eq.70}
\end{eqnarray}
where $m = 1,\:2,\:\ldots$ and $n = 0,\:1,\:\ldots\,$.
\end{proposition}

\begin{proof}
Making use of (\ref{eq.30}), we can show
\begin{equation}
 \Ld{\XX_0} \Omega_0 = 0, \quad
 \Ld{\YY_1} \Omega_0 = 2 \Omega_1
\end{equation}
with a somewhat tedious calculation.
Using (\ref{eq.64}), the equation above and
Leibnitz rule, we can derive Eqs.~(\ref{eq.69})
and (\ref{eq.70}) for the case $m=1$.
Further, we see $\Ld{R^m \XX}\Omega_n = \Ld{\XX}\Omega_{m+n}$
$\forall\XX\in\vect$ as a specific case of (\ref{eq.49}),
since $\ed\Omega_n = 0$ as in (\ref{eq.59}).
\end{proof}

Since $\Omega_n$ are closed 2-forms, the proposition says
\begin{eqnarray}
 \ed\ip{\XX_{m-1}}\Omega_n &=& 0, \label{eq.72} \\
 \ed\ip{\YY_m}\Omega_n &=& (3 - m - n) \Omega_{m+n}.
\end{eqnarray}
Apparently, $\Omega_n$, $n = 1,\:2,\:4,\:5,\:\ldots$ are exact.

Let $\zeta_n$, $n = 0,\:1,\:\ldots$ be the 1-forms
\begin{equation}
 \zeta_n := \ip{\XX_n}\Omega_0 = (\ip{\XX_0}\Omega_0)\circ R^n.
  \label{eq.74}
\end{equation}
As expressed in (\ref{eq.72}), 1-forms $\zeta_n$ are
closed. Since we already know the Lie derivative of $\XX_n$ and
$\Omega_0$ in the direction of $\XX_{m-1}$ and $\YY_m$, we can
easily verify that
\begin{eqnarray}
 \Ld{\XX_{m-1}}\zeta_n &=& 0, \\
 \Ld{\YY_m}\zeta_n &=& (1 - m - n) \zeta_{m+n}, \label{eq.76}
\end{eqnarray}
where $m = 1,\:2,\:\ldots$ and $n = 0,\:1,\:\ldots\,$.

\begin{theorem}
The 1-forms $\zeta_n$, $n = 0,\:2,\:3,\:\ldots$ defined by
(\ref{eq.74}) are exact and are written as
$\zeta_0 = \ed I_0$, $I_0 := \length$ and
\begin{equation}
 \zeta_n = \ed I_n, \quad
 I_n := \frac{1}{1-n}\, \zeta_{n-1}(\YY_1),
 \quad n = 2,\:3,\:\ldots\,.  \label{eq.77}
\end{equation}
The functionals $I_n$, $n = 0,\:2,\:3,\:\ldots$ are in involution
with respect to the Poisson bracket associated with $H_k$ with
arbitrary $k$.
The vector fields $\XX_n$, $n = 0,\:2,\:3,\:\ldots$ of (\ref{eq.61})
are Hamiltonian vector fields of $I_n$ with respect to $H_0$, namely,
$\XX_n = J \grad I_n$.
\end{theorem}

\begin{proof}
Since $\zeta_n$ are closed 1-forms, (\ref{eq.76}) can be
written as $\ed\ip{\YY_m}\zeta_n = (1 - m - n) \zeta_{m+n}$, from
which we see that $\zeta_n$, $n = 2,\:3,\:\ldots$ are exact
1-forms written as in (\ref{eq.77}). The same is easy
for the case $n=0$.
Using (\ref{eq.55}) and (\ref{eq.58}), we see
\begin{eqnarray*}
 \Poisson{I_i}{I_j}{}_{H_k}
 &=& \ed I_j (H_k \ed I_i) = \zeta_j (H_k \zeta_i)
  = \Omega_0(\XX_j,\:H_k \zeta_i) \\
 &=& -\zeta_i(R^k \XX_j) = - \Omega_{i+j+k}(\XX_0,\:\XX_0) = 0,
\end{eqnarray*}
namely, the functionals $I_n$ are in involution
with respect to the Poisson bracket associated with $H_k$.
Using (\ref{eq.57}), we see
\[
 \Omega_0(J\grad I_n,\:\YY) = \Riemann{\grad I_n}{\YY}
  = \ed I_n(\YY) = \zeta_n(\YY) = \Omega_0(\XX_n,\:\YY)
\]
for all $\YY\in\vect$. This implies $\XX_n = J\grad I_n$.
\end{proof}
Although we gave it in the proof, an explanation for the
statements of the theorem other than $\zeta_n = \ed I_n$ can
be found in the general theory \cite{Ma,G-D}.

The missing piece, functional $I_1$
such that $\XX_1 = J\grad I_1$ does not exist within $\func$
(the possible candidate for $I_1$ in other treatments
is the total torsion \cite{L-P,Sa}).

The theorem above proves the inspection of Langer and Perline
\cite{L-P} saying
\begin{equation}
 I_n = \frac{1}{n-1}\int_{-\infty}^\infty \ed s\,
  \ints \inner{\kappa \NN}{\XX_{n-1}}.
\end{equation}
Partly, the same is given in \cite{Sa}. The proof is as follows:
Inserting $\ders s = 1$ into the integrand and then making partial
integration, we see that the right-hand side of the equation above
equals to
\[
  \frac{1}{n-1}\int_{-\infty}^\infty \ed s\,
           \inner{- s \kappa \NN}{\XX_{n-1}} \\
  = \frac{1}{n-1}\, \Riemann{J^{-1}(s \kappa \BB)}{\XX_{n-1}}
  = \frac{1}{n-1}\, \Omega_0(\YY_1,\:\XX_{n-1}),
\]
which obviously coinides with (\ref{eq.77}).
The surface term in this partial integration is absent, since
$\ints \inner{\kappa \NN}{\XX_{n_1}}\in\alge{2}$
follows from $\inner{\kappa \NN}{\XX_{n_1}}\in\alge{2}$ by
virtue of the fact \cite{L-P} that
$\ints \inner{\kappa \NN}{\XX_{n-1}}$ can be written as
polynomials in $\ders{}^n\kappa$ and $\ders{}^n\tau$.

\end{document}